# A new approach in the detection of weak γ-ray peak of the radioactive waste in tomography γ scanning

Zhang Jinzhao[1](张金钊), Tuo Xianguo[1,2,3](庹先国)

*(1) Chengdu University of Technology Applied Nuclear Techniques in Geoscience Key Laboratory of Sichuan Province, Chengdu 610059, China*

*(2) Chengdu University of Technology State Key Laboratory of Geohazard Prevention & Geoenvironmental Protection, Chengdu 610059, China*

*(3) Southwest University of Science and Technology, Mianyang 621010, China*

**Abstract:** We demonstrate a new approach to efficiently detect weak γ-ray peak of the radioactive waste in tomographic γ scanning (TGS). In the TGS measurement, γ-ray peak identification is usually difficult due to the short measurement time that results in a lower γ-ray energy produced by the decay. Consequently, the resulting significant scattering in the low-energy side leads to strong statistical fluctuations and low detection efficiency that overwhelm the γ-ray peak. Here, we propose the use of shift invariance wavelet algorithm for low-energy part of the spectrum for weak γ-ray peak smoothing. The proposed algorithm not only overcomes the pseudo-Gibbs in the high-resolution γ-ray spectrum de-noising by the traditional wavelet transform, but also keeps quality of the weak γ-ray characteristic peak as well. Our new approach shows a significantly improved performance of the figure of merit (FOM) together with lower limit of quantitation (LLOQ) compared with the traditional wavelet transform. These results indicate the significant potential of our approach that could effectively increase the accuracy of weak γ-ray detecting activation analysis in TGS.

## 1 Introduction

The Tomographic Gamma Scanning (TGS) technique is a relatively new method in the field of nondestructive assay (NDA) of radioactive waste detection [1, 2]. It is used in industrial CT imaging technology to solve the problem of inaccurate attenuation correction that involves the uneven distribution of the sample medium. Thus, it improves the accuracy of the content of the non-uniform analysis of radioactive samples in the γ-ray spectroscopy measurements. TGS technology is currently the preferred technology for nuclear waste NDA detection [3-5].

In the TGS measurement, since the level of radionuclides to be analyzed is lower, the measurement time is shorter and the resulting lower γ-ray energy produced by the decay and scattering in the low-energy side is more serious. This leads to the higher background count rate in actual measurement of low-energy γ spectrum, and the characteristic peaks get overwhelmed by its statistical fluctuations. Hence, it is difficult to assess γ-ray peak identification, peak area calculation and confidence intervals.

To improve the system's ability to detect weak γ peak, the general approach is to increase the volume of the sample to improve radioactivity of the sample. This subsequently improves the detection efficiency peaks of large area detectors using anti-coincidence techniques. However, it is difficult to implement these methods in the actual measurement of TGS. In addition, smoothing algorithm is needed to deal with the actual measured spectrum to improve weak γ peak detection capabilities.

Least squares smoothing method have good effects in the treatment of γ energy spectrum with low energy resolution using a NaI detector. However, for more frequently changing γ spectrum data with high resolution, it can filter out statistical fluctuations along

Supported by National Natural Science Foundation of China (41274109, 41025015), Scientific and Technological Innovative Team in Sichuan Province (2011JTD0013) and '863' Program of China (2012AA063501)
E-mail:zhangjinzhao_cdut@163.com

with some high-frequency signals, which may lead to γ spectrum distortion. This implies that the algorithm is not conducive to extract weak γ peaks in the high scattering background. Wavelet transform has a better effect in processing low-resolution γ spectrum, and can effectively extract the weak γ peak information[6, 8]. But the neighborhood of discontinuities will exhibit visual artifacts in processing high-resolution γ spectrum. We attribute some of these limitations to the lack of translation invariance of the wavelet basis. In this paper, we propose the shift-invariant wavelet smoothing method of TGS spectrum. The proposed method maintains the measurement spectrum quality and also improves performance of the figure of merit (FOM) and lower limit of quantitation (LLOQ), which is the basic data of radionuclide identification and quantification.

## 2 Theoretical basis

### 2.1 Translation Invariant Wavelet Theory

Shift-invariant wavelet transform is an algorithm based on wavelet transform. In the last few years, there has been considerable interest in the use of wavelet transforms for removing noise from signals. Wavelet transform theory is described as follows: signals are projected to subspace with different frequencies and then processed in frequency space for signal reconstruction. Energy normalized wavelet family is obtained by basic wavelet function through stretching and translation:

$$\psi_{\tau,a}(t) = \frac{1}{\sqrt{a}} \psi(\frac{t-\tau}{a}) \qquad (a>0, \tau \in R).$$

$\psi_{\tau,a}(t)$ is a self-similar function family by scale $a$ extension transformation and time $\tau$ translation transformation of basic wavelet $\psi(t)$. We adopted this set of functions to analyze signal decomposition[7-9].

Translation corresponds to non-uniform sampling. When the scale displacement sampling interval is exponentially doubled, orthogonal wavelet function set cannot be obtained from the perspective of multi-scale matching for the local structure of the signal characteristics. Therefore, the wavelet transform sometimes exhibits visual artifacts in the neighborhood of discontinuities[10-12].

The steps of shift-invariant wavelet transform de-noising are described as follows:

1. Shift

Firstly, time-domain translation operator is introduced. For signal $y_t$ $(0 \leq t \leq N)$, $S_h$ is defined as a shift operator with cyclic shift amount of $h$.

$$S_h(x_t) = x(t+h) \bmod N \ ;$$

2. Smoothing

We select the 'sym' series sym8 as the wavelets in this study. Secondly, we select the decomposition scale. Different frequency bands correspond to different scale decomposition. We then determine the γ spectrum signal decomposition of five layers by which we analyze the spectrum. Next, we select the threshold as follows:

$$\lambda = \sigma \sqrt{2\lg(M)} \left(1 + \frac{1}{j}\right)$$

$\lambda$ is the threshold;

$M$ is the length of γ energy spectroscopy data;

$$\sigma = \frac{median(|w_{j,k}|)}{0.6745};$$

Lastly, the spectrum signal corresponding to the high and low frequency coefficients which deal with the threshold is decomposed.

3. Opposite translation

The inverse shift spectrum is smoothed by the wavelet:

$$S-h(x_t) = x(t-h) \bmod N$$

4. Average

Artificial oscillation amplitude is minimized by selecting the optimal translation parameters h. However, when the spectrum contains a plurality of singular points, it is possible that a singular point shift amount is optimal, and the singular point to another shift amount is the worst. Therefore, for a complex spectrum, a range of translation was circulation translation operated, and then the result was averaged in order to eliminate the phenomenon of oscillation.

## 2.2 Evaluation of performance of weak γ peak detection

Figure of merit and lower limit of quantitation are usually used as the criteria to evaluate the device detection capabilities of weak γ peaks. Figure of merit is a function of the smallest measuring radioactivity counting time T, which corresponds to the least relative standard deviation. The bigger the figure of merit, the less time it takes to measure the unit of activity. On the contrary, the smaller the quality factor, the longer is the time required for measurement.

$$Q = v^2 n_0^2 / \left(\sqrt{n_0 + n_b} + \sqrt{n_b}\right)^2$$

$n_0$ is the sample net peak area;

$n_b$ is the background net peak area (this is continuum counts )

$v$ is the relative standard error of $n_0$;

We considered 95% confidence probability assuming the net area greater than the determining limit. We note that the sample is radioactive when the net area greater than the detection limit. This means that this device can be measured to a minimum radioactivity. Quantitative measurements of the standard error of the net count can be considered less than a predetermined value (generally take 10%). This corresponds to the count value of the net that can be accurately recognized with minimum radioactivity that we call lower limit of quantitation.

$$L_c = K_\alpha \sqrt{2n_b}$$

$L_c$ is the function of determine the limit.

The $K_\alpha = K_\beta = K = 1.645$ is taken by 95% confidence probability.

$$L_D = K^2 + 2L_c$$

$L_D$ is the function of detection limit.

$$L_Q = 50\left[1 + \left(1.0 + 0.08 \cdot N_b\right)^{\frac{1}{2}}\right]$$

$L_Q$ is the function of lower limit of quantitation.

We only calculate the lower limit of quantitation.

## 3 TGS Measurement Experiment

The TGS mechanism developed by our group at Chengdu University of technology consists of the modules: the level of the mobile/rotation platform, lifting platform detectors, radioactive lifting platform and a transmission source shield. A picture of the system is shown in Fig. 1. Automation of level/rotation, vertical and rotational platform is controlled by a Process Logic Controller (PLC). The system consisted of a GEM50P4-83 detector which produced by ORTEC and a 10mCi $^{152}$Eu transmission source. Fifty two γ rays were product by $^{152}$Eu. As shown in Tab. 1, only 12 rays were calculated with relatively large fraction.

Tab. 1 Ray fraction and energy of photon emission products: $^{152}$Eu

| Fraction | Energy(keV) | Fraction | Energy(keV) |
|---|---|---|---|
| 0.013805 | 1212.8 | 0.12741 | 778.89 |
| 0.022144 | 411.11 | 0.13302 | 1112.00 |
| 0.028114 | 443.98 | 0.14441 | 964.01 |
| 0.041601 | 867.32 | 0.20747 | 1085.80 |
| 0.074935 | 244.69 | 0.26488 | 344.27 |
| 0.099630 | 1085.80 | 0.28432 | 121.78 |

The detector was housed inside a cylindrical shield of 50 mm annular thicknesses. The collimator of detector and transmission source used was made of lead with an aperture of 87.5mm and 10mm diameter. The data acquisition and analysis software platform consisted of Canberra's GENIE-2000. Measuring waste drums filled with Acrylonitrile Butadiene Styrene plastics (ABS) was the national standard 200-L waste drums. Drum wall thickness consisted of 1.25 mm with the sample volume: 5 cm×5 cm ×5 cm; density: 1.07 g/cm$^3$.

Experimental procedure was divided into three steps. First of all the background spectrum was collected by the HPGe detector in the lead house. The transmission source was then exposed to the detector and an un-attenuated spectrum was collected. Finally, a drum containing radioactive sources and ABS were loaded on to the rotating platform.

Fig. 2 is the TGS system schematic diagram and modules geometry.

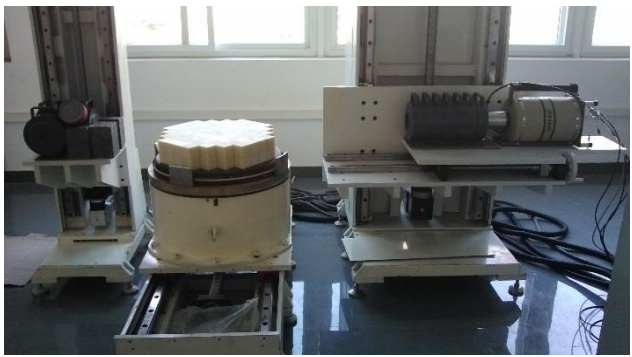

Fig. 1 Tomographic Gamma Scanner.

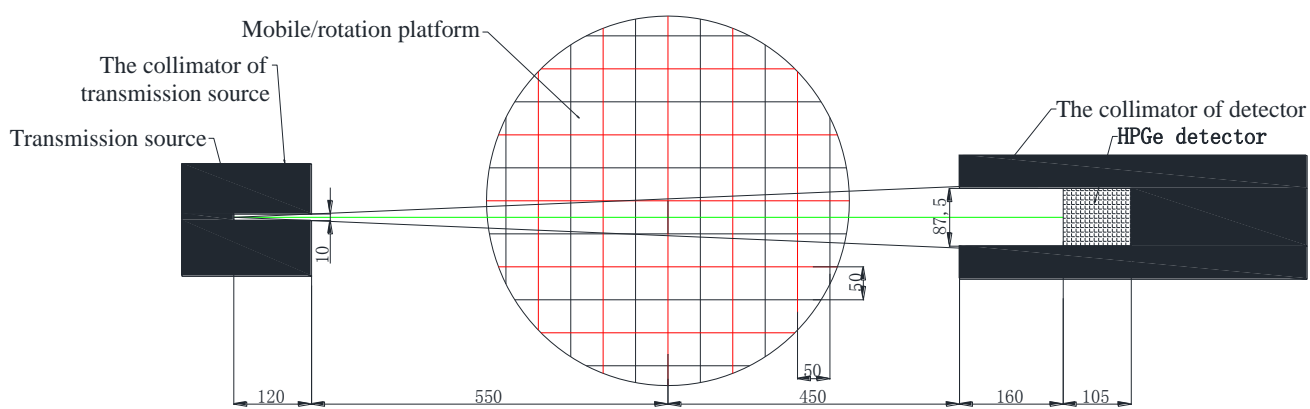

Fig. 2 A schematic diagram of the TGS system (mm).

## 4   Analysis and discussion

We note that lower environment background is very conducive for weak γ peak extraction in the low γ radioactivity measurements. Therefore, the detector of TGS was housed inside a cylindrical shield to reduce γ radiation background.

Fig. 3 show the original $^{152}$Eu energy spectrum and the environmental background. Fig. 4 shows the attenuated $^{152}$Eu energy spectrum and the environmental background. As shown in Fig.3-4, we observe that the count rate of environmental background is far less than the count rate spectrum of $^{152}$Eu. This implies a greater source activity, and thus in the actual measurement environmental background can be ignored.

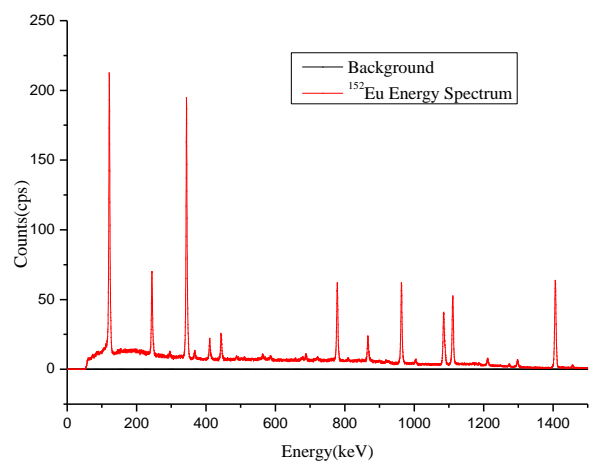

Fig. 3 $^{152}$Eu and background energy spectrum.

After adding the sample, low energy γ ray was observed to be attenuated faster and the counting rate dropped significantly indicating low-energy platform forming the weak peaks of lower count rate. It is to be noted that Compton scattering causes more counts rate of low energy part. Hence, its statistical fluctuation influences on weak γ peak that leads to counts more than the environment background. So, TGS system can detect weak γ peak with a high Compton scattering platform.

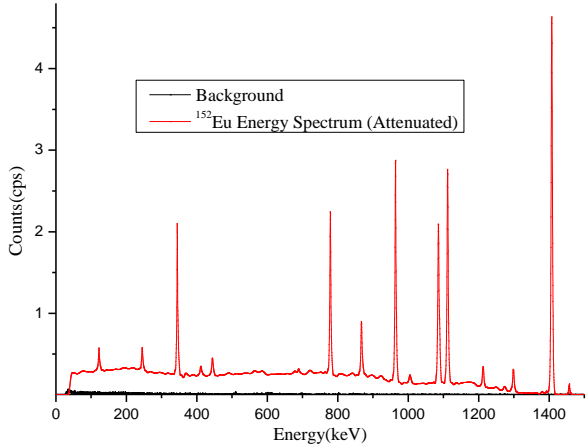

Fig. 4 [152]Eu and background energy spectrum (attenuated).

The γ energy spectrum was smoothed to improve the γ energy spectrum analysis capability and to reduce the actual measuring spectrum statistical fluctuation by least squares filtering, wavelet transform and shift-invariance wavelet respectively. Weak γ peak detection capability was evaluated by net peak area, figure of merit and lower limit of quantitation.

Five-point and eleven-point least squares filtering formulas are described as follows:

$$\bar{Y}_i = \frac{1}{35}(-8y_{i-2} + 12y_{i-1} + 17y_i + 12y_{i+1} - 8y_{i+2})$$

$$\bar{Y}_i = \frac{1}{429}(-36y_{i-5} + 9y_{i-4} + 44y_{i-3} + 69y_{i-2} + 84y_{i-1} + 89y_i + 84y_{i+1} + 69y_{i+2} + 44y_{i+3} + 9y_{i+4} - 36y_{i+5})$$

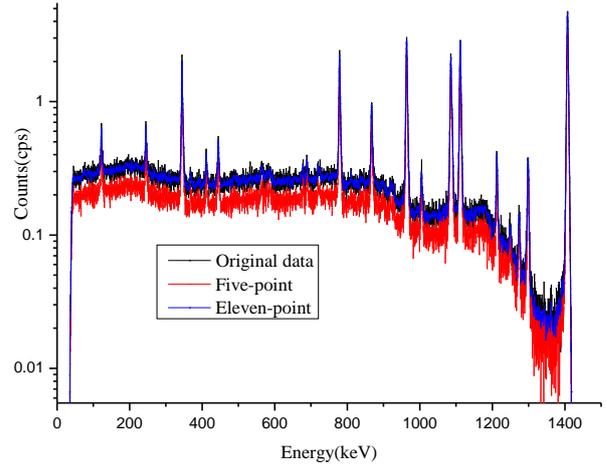

Fig. 5 Five point and eleven point least squares smoothing.

As shown in Fig. 5, after five-point least squares filtering smoothing, the spectrum count rate is smaller than the original data. As can be seen in the figure, the entire spectrum of statistical fluctuation is still very high. Further, for eleven-point least squares filtering smoothing, spectrum deformation is smaller with almost unchanged scattering background and characteristic peak of the spectrum. However, statistical fluctuation is still observed to be high. This implies that the smoothing effect of least squares filtering is not obvious.

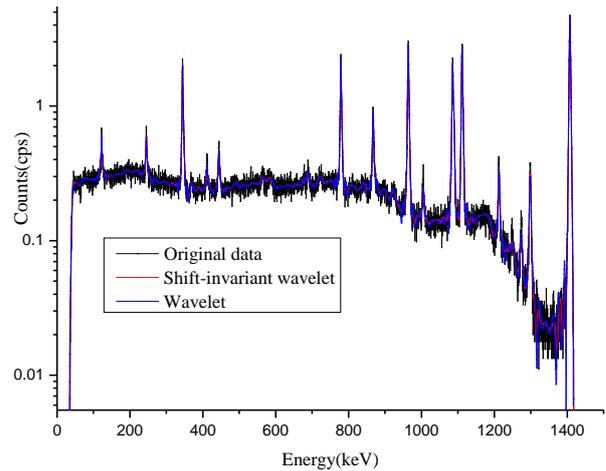

Fig. 6 Wavelet and shift-invariant wavelet transform smoothing.

As shown in Fig. 6, after wavelet transforms smoothing, wavelet transform performs better than the least squares filtering. However, viewing from the sides of the characteristic peaks, obvious fluctuation phenomena were observed that affected the full spectrum of smoothing effect. Using shift-invariant wavelet transform, the small peaks of low-energy part are very smooth and the smoothing effect is significant. Thus, this approach is conducive to a weak peak detection and peak area calculation.

Tab. 2 The peaks of $^{152}$Eu figure of merit

| Energy (keV) | Original data | S-I wavelet | Wavelet |
|---|---|---|---|
| 121.78 | 1.55E-02 | 2.38E-02 | 1.57E-02 |
| 244.69 | 2.29E-02 | 2.49E-02 | 1.71E-02 |
| 344.27 | 6.38E-01 | 7.80E-01 | 6.99E-01 |
| 411.11 | 6.63E-03 | 8.44E-03 | 4.38E-03 |
| 443.98 | 2.24E-02 | 2.68E-02 | 2.52E-02 |
| 778.89 | 1.02E+00 | 1.14E+00 | 4.47E-01 |
| 867.32 | 1.40E-01 | 1.68E-01 | 1.15E-01 |
| 964.01 | 1.99E+00 | 2.19E+00 | 1.22E+00 |
| 1085.78 | 1.87E+00 | 1.70E+00 | 7.64E-01 |
| 1112.02 | 2.24E+00 | 2.35E+00 | 1.37E+00 |
| 1212.80 | 3.06E-02 | 2.94E-02 | 8.29E-03 |
| 1407.95 | 8.37E+00 | 8.59E+00 | 8.52E+00 |

As shown in

Tab. 2, we demonstrate that the shift-invariant wavelet transform figure of merit show a significant increase compared with original data and wavelet transform. We observe that the wavelet transform figure of merit has obvious fluctuation with the greater area characteristic peaks figure of merit better than the original data, and the smaller area characteristic peaks figure of merit worse than the original data. The main reason is that wavelet transform neighborhood of discontinuities exhibit visual artifacts, which can affect its smoothing effect. For shift-invariant wavelet transform, $^{152}$Eu characteristic peaks figure of merit has a greater value. Since it overcomes the disadvantages of the wavelet transform, the high peaks and low count rate peaks at count rates produce better results.

Tab. 3 The peaks of $^{152}$Eu ratio of net peak area and lower limit of quantification.

| Energy (keV) | Original data | S-I wavelet | Wavelet |
|---|---|---|---|
| 121.78 | 2.89E-02 | 3.16E-02 | 2.63E-02 |
| 244.69 | 3.19E-02 | 3.29E-02 | 2.73E-02 |
| 344.27 | 1.77E-01 | 1.91E-01 | 1.83E-01 |
| 411.11 | 1.85E-02 | 1.89E-02 | 1.50E-02 |
| 443.98 | 3.23E-02 | 3.37E-02 | 3.37E-02 |
| 778.89 | 2.21E-01 | 2.23E-01 | 2.11E-01 |
| 867.32 | 8.03E-02 | 8.37E-02 | 7.46E-02 |
| 964.01 | 3.19E-01 | 3.28E-01 | 3.31E-01 |
| 1085.78 | 2.52E-01 | 2.91E-01 | 2.79E-01 |
| 1112.02 | 3.41E-01 | 3.41E-01 | 3.42E-01 |
| 1212.80 | 4.18E-02 | 4.03E-02 | 3.71E-02 |
| 1407.95 | 6.71E-01 | 7.11E-01 | 7.13E-01 |

The ratio of net peak area and the lower limit of quantification is shown in Tab. 3. The lower limit of quantification single cannot fully show the ability of the algorithm to improve the system of weak gamma peak detection. It only shows the ability to reduce the background count. As observed in Table 3, the ratio of lower limit of quantification and net peak area show the ability to improve the detection of weak peaks. The greater the data, the stronger is the ability for weak peak detection. It illustrates that lower limit of quantification of γ energy spectrum that is smootheds by shift-invariant wavelet transform is much better than the traditional wavelet transform, especially in the higher side of the low-energy scattering case. So, it can be stated that algorithm is more suitable for smoothing γ energy spectrum data to improve the detection ability of weak peaks.

## 5 Conclusion

We have presented a new approach to detect weak γ-ray peak in tomographic gamma scanning (TGS). It is based on preprocessing of the γ spectrum as measured by the TGS system and using the shift-invariant wavelet transform. We have shown that not only small calculation errors in the peak area in both the low-energy part and the high-energy part of the spectrum can be achieved, but also significantly improved quality factor and lower limit of quantification of the original energy spectrum can be obtained that simultaneously reduce statistical fluctuations compared with the least squares filtering and traditional wavelet based methods. Due to the limitations of the total measurement time, the measured γ spectrum of TGS requires smaller statistical fluctuation with higher figure of merit and lower limit of quantification. These three important components for efficiently measuring spectra can be improved by employing shift-invariant wavelet transform. We envision that the application of this method has the potential to overcome the limitations in the existing approaches that could significantly improve the ability to detect a weak γ peak.

# 6  Acknowledgements

The authors wish to acknowledge the support of the Applied Nuclear Techniques in Geoscience Key and State Key Laboratory of Geohazard Prevention & Geoenvironmental Protection, Chengdu University of. Technology.